\documentclass[aps,twocolumn,floatfix]{revtex4}
\usepackage{graphicx}

\begin{document}

\title{Testing Strong-field Gravity with Quasi-Periodic Oscillations}
\author{Simon DeDeo}
\affiliation{Department of Astrophysical Sciences, 
Princeton University, Princeton, New Jersey, 08544}
\author{Dimitrios Psaltis}
\affiliation{Physics and Astronomy Departments, University of Arizona,
1118 E. 4th St., Tucson, AZ 85721}

\begin{abstract}
The frequencies of quasi-periodic oscillations around neutron stars
are believed to be related to characteristic frequencies in the
gravitational fields of the compact objects. In different variability
models, these include the Keplerian, epicyclic, and Lense-Thirring
frequencies, which depend mostly on the properties of the stellar
spacetimes.  We argue that quasi-periodic oscillations in the X-ray
flux of neutron stars can be used to map the external spacetimes of
the compact objects and, therefore, lead to direct tests of general
relativity in the strong-field regime. In particular, we show that
particular extensions of General Relativity, in which the
gravitational force felt by matter is mediated by both a rank-two
tensor and a scalar field, can be constrained by current observations.
\end{abstract}

\maketitle

\section{Introduction}

Neutron star properties, such as radius and mass, help probe the
nature of strong-field gravity. In a previous paper~\cite{dp2}, we
showed how the measurement of the redshift of atomic lines emitted
from the surface of a neutron star can be a powerful tool for
investigating the properties of gravity in the strong-field regime. In
particular, we found that our current lack of knowledge about the
behavior of gravity in the strong-field regime far outweigh our
current lack of knowledge about the neutron star equation of state.

In this paper, we expand that investigation to show how the properties
of the spacetime around a neutron star are at least as sensitive to the
choice of gravity theory in the range currently allowed by weak-field
observations as they are to the exact nature of the neutron star
equation of state. As in~\cite{dp2}, we choose a particular
single-parameter scalar-tensor theory but this time we extend our
consideration to the spacetime outside the neutron star surface.

In particular, we examine the way in which deviations from general
relativity may affect the frequencies of quasi-periodic oscillations
(QPOs) found in many X-ray binaries. QPOs are of contested origin; we
refer the reader to Refs.~\cite{v00,p01}. All models, however, agree
that the processes responsible for QPO production take place in a
region of order the Schwarzschild radius around the compact
object. This makes QPOs excellent probes of strong-field gravity,
especially if, as many believe, their frequencies are directly related
to the characteristic frequencies of test particles near the innermost
stable circular orbit (ISCO). The growing numbers of QPO observations
promises an ideal testing ground for the predictions of general
relativity.

We provide two new methods of constraining deviations from general
relativity with QPO observations. The first method relies only on
general arguments about the maximum coherent frequency of the
lowest-order modes near the surface of neutron stars~\cite{mlp98}. The second
method, which relies upon a particular interpretation~\cite{svm99} of
a commonly observed QPO phenomenon -- the correlation in peak position
found between high frequency QPO pairs~\cite{pbv98,bpv02} --
approaches the binary pulsar tests in its ability to constrain the
particular scalar-tensor effect considered. The generality of our
first method demonstrates the importance of examining a wide range of
high-energy X-ray phenomena of diverse nature and origin for clues
about the properties of strong-field gravity. The power of our second
method demonstrates how the development of more comprehensive physical
accounts of these phenomena can lead to new advances in our ability to
use the universe as a gravity laboratory, probing spacetime curvatures
many orders of magnitude higher than those previously accessible to
precision gravity tests.

Our paper is organized as follows. In section two, we describe the
field equations and stellar structure equations for stars in a general
scalar-tensor theory and we solve them through a combination of
analytic and computational techniques. In section three, we discuss in
detail the previously unremarked-upon properties of test particle
orbits around such stars and, in section four, we apply these results to
show explicitly how QPO properties can be used to constrain the nature
of strong-field gravity.

\section{Equations of Stellar Structure}

Following~\cite{de92} as an example of a gravity theory that deviates
from GR in the strong-field regime, we consider a theory containing a
scalar field coupled to gravity with the action

\begin{eqnarray}
\nonumber S & = & \frac{1}{16\pi G_{*}}\int  d^4x \sqrt{-g_{*}}(R_{*}-2g^{\mu\nu}_{*}\phi_{,\mu}\phi_{,\nu}) \\
& & + S_m[\Psi_m,A^2(\phi)g_{*\mu\nu}].
\label{lagrangian}
\end{eqnarray}
Here $\Psi_m$ refers collectively to all matter fields other than
$\phi$, $G_*$ is a dimensional constant, and $A(\phi)$ is a function
of $\phi$. The $(0,2)$ tensor $g_{* \mu\nu}$ is the Einstein frame
metric; all matter (apart from $\phi$), flows on geodesics of the
physical (or ``Brans-Dickie'') frame metric, $A^2(\phi)g_{* \mu\nu}$,
which we will refer to as $g_{\mu\nu}$. The field equations for the
scalar field $\phi$ are particularly simple in the Einstein  frame where
$\phi$ appears to be non-minimally coupled to gravity; because of the
unphysical nature of the metric $g_{* \mu\nu}$, however, confusion may
arise; the reader is referred to~\cite{ssn98} for a deeper discussion
of these issues. Unless otherwise specified, all quantities reported
in this paper are measured in the physical frame.

We summarize the field equations, following~\cite{de93}:

\begin{eqnarray}
\nonumber R_{* \mu\nu} & = & 2\partial_{\mu}\phi\partial_{\nu}\phi + 8\pi G_*\left(T_{* \mu\nu}-\frac{1}{2}T_*g_{*\mu\nu}\right), \\
\nonumber \nabla_{*\nu}\nabla_*^{\nu}\phi & = & -4\pi G_*\alpha(\phi)T_*.
\end{eqnarray}
The equation of motion for matter is:

\begin{displaymath}
\nonumber \nabla_{*\nu}T^{\nu}_{*\mu}=\alpha(\phi)T_*\nabla_{*\mu}\phi,
\end{displaymath}
where starred subscripts refer to operations using the Einstein frame
metric, and $\alpha(\phi)=d \ln A(\phi)/d\phi$. The stress energy
tensor in the Einstein frame, $T_*^{\mu\nu}$, is equal to
$A^6(\phi)T^{\mu\nu}$, where $T^{\mu\nu}$ is measured in the physical
frame.

We see that the scalar field, $\phi$, is sourced by the trace of the
stress-energy tensor, proportional to the logarithmic derivative of
the warp factor, $A(\phi)$. Thus, though the cosmological value of
$\phi$ may be zero, a strong enough concentration of matter can
support a non-zero value of $\phi$. Given a sufficiently dense
concentration of matter, and a sufficiently steep $A(\phi)$, a region
of space with non-zero $\phi$ may become energetically favorable.

Solar system experiments constrain the function $A$ to be very flat at
the cosmological value of $\phi$, denoted by $\phi_0$. If, however, the
$\phi$ field inside a compact object fluctuates just far enough away
from cosmological to discover a steep part of $A$, perhaps during a
violent formation event, the system will be able to reach the more
energetically favorable configuration, with large non-zero values for
the scalar field $\phi$ near the center of the object. This is the
phenomenon of ``spontaneous scalarization.''

This phenomenon occurs \emph{in general} for functions $A(\phi)$ when
the derivative $d\ln A/d\phi |_{\phi_0}$ is sufficiently
negative~\cite{de93}, although the precise configuration of the
neutron star and its external spacetime may vary with the particular
choice of $A(\phi)$. We consider only the first term in the Taylor
expansion of $\ln A(\phi)$ that leads to negative curvature,

\begin{equation}
\label{a}
A(\phi)=e^{\frac{1}{2}\beta\phi^2},
\end{equation}
where $\beta$ is a real number. This is the same form as used
in~\cite{de96}, where it is referred to as the ``quadratic model''
because of its relationship to the Taylor expansion of $\ln
A(\phi)$. Other investigations of scalar-tensor theory have chosen
different forms for $A(\phi)$; see~\cite{w02} for a comparative list
of choices of $A(\phi)$, as well as for how to rewrite the action in
Eq.~\ref{lagrangian} explicitly for particular choices of $A(\phi)$
used in the literature, in both the Einstein and physical frame.

Again following~\cite{de93}, we can write the relativistic equations
of stellar structure for a neutron star. We write our metric as:

\begin{eqnarray}
\label{metric}
\nonumber ds_*^2 & = & -e^{\nu(r)}dt^2+\left[1-\frac{2\mu(r)}{r}\right]^{-1}dr^2 \\ & & +r^2(d\theta^2+\sin^2\theta d\phi^2),
\end{eqnarray}
and describe the matter fields as a perfect fluid in the physical frame: 

\begin{displaymath}
\nonumber T_{\mu\nu}=(\rho+p)u_{\mu}u_{\nu}+pg_{\mu\nu}.
\end{displaymath}

The differential equations to be integrated are then:

\begin{eqnarray}
\label{eqs}
\mu^{\prime} & = & 4\pi G_*r^2A^4\rho+\frac{1}{2}r(r-2\mu)\psi^2, \\
\nu^{\prime} & = & 8\pi g_*\frac{r^2A^4p}{r-2\mu}+r\psi^2+\frac{2\mu}{r(r-2\mu)}, \\
\phi^{\prime} & = & \psi, \\
\nonumber \psi^{\prime} & = & 4\pi G_* \frac{rA^4}{r-2\mu}[\alpha(\rho-3p)+r(\rho-p)\psi
\nonumber\\
& &\qquad\qquad\qquad
+\frac{\mu}{r(r-2\mu)}\psi], \\
N^{\prime} & = & 4\pi nA^3r^2(1-\frac{2\mu}{r})^{-1/2}, \\
\nonumber p^{\prime} & = & -(\rho+p)\left[4\pi G_*\frac{r^2A^4p}{r-2\mu}+\frac{1}{2}r\psi^2
   \right.
\nonumber\\
& & \qquad\qquad\qquad\left. +\frac{\mu}{r(r-2\mu)}+\alpha\psi\right]. 
\label{eqs2}
\end{eqnarray}

Here $N$ is the baryon number, $n$ the number density, $\rho$ the
density, and $p$ the pressure. We supplement these equations with a
(zero temperature) equation of state, $p=p(\rho)$ and
$n=n(\rho)$. Primes denote derivatives with respect to coordinate
radius $r$.

Having chosen an equation of state, we then integrate
Eqs.~\ref{eqs}--\ref{eqs2} from the center of the star, where we
specify

\begin{eqnarray*}
\mu(0) = \nu(0) & = & 0, \\
\phi(0) & = & \phi_c, \\
p(0) & = & p_c, \\
N(0) & = & n_c, \\
\rho(0) & = & \rho_c,
\end{eqnarray*}
to the surface, where $p=0$. We then must integrate the equations for
$\nu$, $\phi$, $\psi$, and $\mu$ from the surface of the star to
infinity, to determine the form of the metric exterior to the star.

Throughout this paper we will be interested only in solutions where
$\phi_0$, the cosmological value of $\phi$, is zero [and
$A(\phi_0)=1$, which can be achieved by renormalizing $A$]. We create
our set of solutions to Eqs.~(\ref{eqs})--(\ref{eqs2}) with a simple
shooting method~\cite{NR}. We note that $\mu$ \emph{does not} have
compact support; outside of the star the scalar field $\phi$ is
non-zero (though relaxing as $1/r$ to zero) and contributing to the
physical Arnowitt-Deser-Misner (ADM) energy felt by an observer far
away. In our terminology, the physical ADM energy is that that would
be inferred, e.g., from test particle orbits near infinity, and is
identical to the $M_{\textrm{\small{ADM}}}$ in~\cite{ssn98}. When
considering solutions to the metric, we renormalize the lapse
function, $\exp{[\nu(r)]}$ so that the physical metric,
Eq.~(\ref{metric}), goes to $\eta_{\mu\nu}$, the Minkowski value, at
infinity.

In an earlier paper \cite{dp2}, we considered three commonly
used equations of state, which cover a broad subset of the wide range
discussed in Cook et al.~\cite{cst94}.  In order of increasing
stiffness, these were EOS~A~\cite{eosa}, EOS~UU~\cite{eosuu}, and
EOS~L~\cite{eosl}. Neutron star models computed with these three
equations of state bracket the uncertainty introduced by our inability
to calculate from first principles the properties of ultra-dense
matter, when condensates or unconfined u-d-s quark matter is not taken
into account.

For this paper, we will use the ``median'' EOS~UU to illustrate the
qualitative behavior of the variables we wish to examine; the results
calculated for UU differ only slightly from those calculated with the
two bracketing equations of state, A and L, in accord with the results
presented in~\cite{dp2}.

For simplicity, we neglect the effect of rotation on the frequencies of orbits in the
external spacetime. In General Relativity, the effect of rotation on
the orbital frequencies is less than 20\% for the inferred spin
frequencies of the neutron stars that show QPOs. We expect rotation to
affect the conclusions in this paper at the same level.

\section{Properties of Scalar Stars}

Our gravity theory is described by the action in
Eq.~(\ref{lagrangian}) and the definition of the function $A(\phi)$ in
Eq.~(\ref{a}). The function $A(\phi)$ has one free, real parameter,
$\beta$, which characterizes the strength of the scalar field coupling
to gravity.

\begin{figure}
\includegraphics[width=8cm]{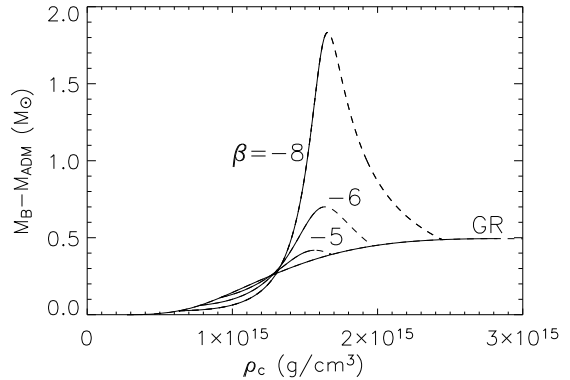}
\caption{Binding energies of our neutron star configurations, as a function of central density, for different values of the parameter $\beta$, for equation of state UU. Dashed lines indicate where a star is unstable to radial perturbations.}
\label{3_1}
\end{figure}

Spontaneous scalarization occurs only when $\beta\leq-4.85$. In
Fig.~\ref{3_1}, we plot the binding energy,
$(Nm_b-M_{\textrm{\small{ADM}}}$, where $m_b$ is the baryon rest
mass), as a function of central density $\rho_c$, using equation of
state UU, for a range of values for the parameter $\beta$. We draw the
reader's attention to the very high binding energies that can be
achieved when $\beta$ is sufficiently large. This large energy is due
mostly to energy released when $\phi$ drops away from its cosmological
value; when a scalarized star is produced, this energy goes into
producing a monopole ``gravitational wave'' in $\phi$.

After the ``turnover point'', \emph{i.e.}, the central density for
which the mass for a particular choice of $\beta$ is at its maximum,
the stars are unstable to radial perturbation and will collapse; these
stars are marked with a dashed line in Fig.~\ref{3_1}. Following the
stability analysis of~\cite{h98}, we know also that when the binding
energy for a scalarized star is greater than that for its
unscalarized, General Relativistic counterpart with the same central
density, \emph{only} the scalarized configuration is
stable. Scalarization is mandatory when the scalarized star has
greater binding energy.

\begin{figure}
\includegraphics[width=8cm]{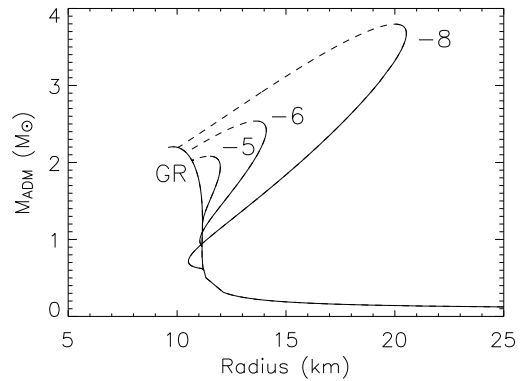}
\caption{The mass-radius relation for our neutron star configurations, for different values of the parameter $\beta$, for equation of state UU. Dashed lines indicate where a star is unstable to radial perturbations. A fiducial value comparable to the most accurately 
measured neutron-star masses is drawn as a horizontal heavy dashed line.}
\label{3_3}
\end{figure}

Fig.~\ref{3_3} shows the mass-radius relationship for the same range
of the parameter $\beta$ and the same equation of state, UU, as
above. Immediately apparent is the result~\cite{ssn98,de93} that
scalarized stars can achieve large masses and large radii.

\section{Orbits in the Vicinity of Scalar Stars}

Here we consider the properties of a test particle in a circular orbit
outside the star. We first consider the orbital frequency of the
particle as seen at infinity. Paralleling the analysis for circular
orbits in the Schwarzschild metric~\cite{MTW}, we find:

\begin{eqnarray}
\nonumber \left(\frac{dr}{d\tau}\right)^2 & = & \frac{\left(1-2\mu/r\right)}{A^2}\left[\frac{\tilde{E}^2}{A(\phi)^2e^{\nu}}-\left(1-\frac{\tilde{l}^2}{A(\phi)^2r^2}\right)\right],
\end{eqnarray}
where $\tilde{E}$ and $\tilde{l}$ are the energy and angular momentum
per unit mass of the test particle, respectively, $\tau$ is the proper
time of the particle, and $r$ is the orbital radius. We note that, in
contrast to the General Relativistic case, $\mu$ is a function of
radius (as is, necessarily, $A$, which is a function of $\phi$). For
the orbit to be circular, we must have

\begin{eqnarray*}
\left(\frac{dr}{d\tau}\right)^2 & = & 0, \\
\frac{d}{dr}\left(\frac{dr}{d\tau}\right)^2 & = & 0.
\end{eqnarray*}
Using these conditions to eliminate $\tilde{E}$ and $\tilde{l}$, and using the lapse function $dt/d\tau$, we find the orbital frequency of a test particle, as measured at infinity, to be

\begin{eqnarray*}
\omega_o & = & \frac{d\phi}{dt}=e^{\nu/2}\sqrt{\frac{2\beta\phi\phi^{\prime}+\nu^{\prime}}{2\beta\phi\phi^{\prime} r^2+2r}} \\
& = & \sqrt{ \frac{\mu}{r^3}\left[\frac{e^{\nu}(1-2\mu/r)^{-1}+\psi(\beta\phi+r\psi/2)r^2e^{\nu}/\mu}{1+\psi r \beta\phi}\right]},
\end{eqnarray*}
where, as before, primes denote derivative with respect to $r$. The
final form of this equation demonstrates how the orbital frequency
approaches the general relativistic value as the scalar coupling,
$\beta$, is taken to zero. In GR, $\psi=\phi=0$ and
$e^{\nu}(1-2\mu/r)^{-1}=1$, so that we obtain the Kepler frequency
$\omega^2=\mu/r^3$.

\begin{figure}
\includegraphics[width=8cm]{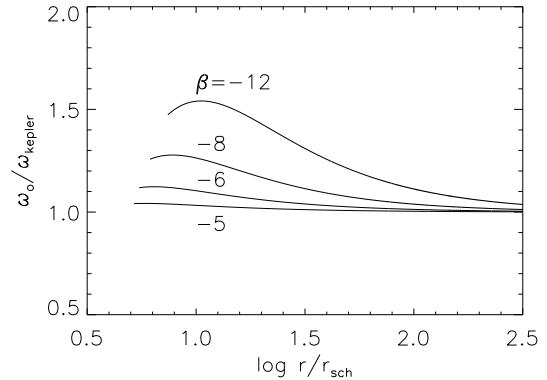}
\caption{The ratio of orbital to ``Keplerian'' frequencies for a star with 
$M_{\textrm{\small{ADM}}}=1.5 M_{\odot}$ for a range of values of
$\beta$, using equation of state UU. Not all the orbits for which
frequencies are plotted here are stable; see Fig.~\ref{4_2}.}
\label{4_1}
\end{figure}

Figure~\ref{4_1} shows the ratio of orbital frequency to the
``Keplerian'' value as a function of the logarithm of the distance
from the center of mass, for a star with $M_{\textrm{\small{ADM}}}=1.5
M_{\odot}$, for equation of state UU and for a range of values of the
parameter $\beta$. We define the Keplerian frequency at radius $r$ to
be the orbital frequency that would be produced at radius $r$ were the
scalarized star replaced by a General Relativistic star with
$M_{\textrm{\small{ADM}}}$ equal to $\mu(r)$. We remind the reader
that $\mu(r)$ is increasing even outside the star, since $\phi$ is
non-zero near the surface.

Comparison of figures~\ref{4_1} and~\ref{3_3} reveals an interesting
and counter-intuitive effect of spontaneous scalarization. Scalarized
stars have radii much greater than their General Relativistic
counterparts with the same $M_{\textrm{\small{ADM}}}$. One would thus
presume that their surface gravity, and thus their orbital frequencies
near the surface, would be lower than the General Relativistic
case. In fact, the case is quite the opposite; near the surface of a
scalarized star the orbital frequency can be up to 50\% higher than
they would be at the same distance for a General Relativistic star
with the same enclosed mass. At distances greater than $\approx 300$
Schwarzschild radii, however, the orbits converge rapidly to their
General Relativistic values. This is a dramatic illustration of how
large strong-field effects can effectively hide themselves in a small
region.

To investigate the stability of these orbits to radial perturbations
we calculate the epicyclic frequency $\kappa$. We find

\begin{displaymath}
\label{epi}
\nonumber \kappa^2=\frac{1}{2}\left(\frac{d\tau}{dt}\right)\frac{d^2}{dr^2}\left(\frac{dr}{d\tau}\right)^2.
\end{displaymath}
The orbital stability criterion is simply that $\kappa^2>0$. The form
of the right hand side of this equation is complicated, but it depends
only on the variables in Eqs.~(\ref{eqs})--(\ref{eqs2}) and their
derivatives.

\begin{figure}
\includegraphics[width=8cm]{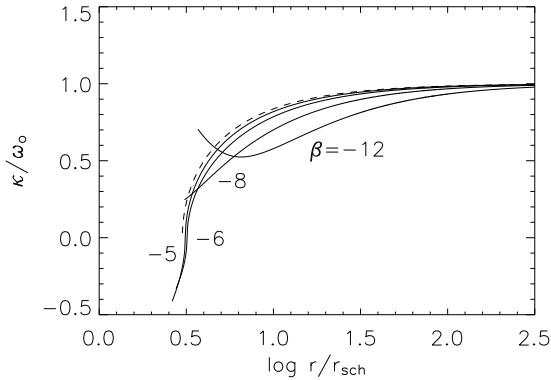}
\caption{The ratio of epicyclic to orbital frequencies for a range of values of $\beta$ and $M_{\textrm{\small{ADM}}}=1.5 M_{\odot}$, using equation of state UU. A negative epicyclic frequency is used here to indicate instability (i.e., $\kappa^2<0$.) The limiting GR case is plotted as a dashed line.}
\label{4_2}
\end{figure}

Figure~\ref{4_2} shows the behavior of the epicyclic frequency as a
function of the logarithm of radius, again with
$M_{\textrm{\small{ADM}}}=1.5 M_{\odot}$, equation of state UU, and a
range of values for the parameter $\beta$. A negative epicyclic
frequency is used here to indicate an imaginary $\kappa$ (i.e.,
$\kappa^2<0$, instability.) We see again how the properties of the
scalarized stars are altered by factors of order unity compared to the
General Relativistic case. The point of instability is changed such
that, as the parameter $\beta$ is made more negative, the ISCO radius
is pushed closer to the surface of the star compared to the GR ISCO
radius. As expected, for radii much larger than the ISCO or stellar
radius, the ratio of epicyclic to orbital frequency converges to the
General Relativistic value.

As can be seen in Fig.~\ref{4_2}, for values of the parameter
$\beta\lesssim -8$ and neutron star masses
$M_{\textrm{\small{ADM}}}\sim1.5 M_{\odot}$, \emph{all} orbits down to
the stellar surface are stable.

\section{Constraints on $\beta$ from Quasi-periodic Oscillations}

The predicted differences in orbital and epicyclic frequencies between
the General Relativistic and scalar-tensor theories discussed in the
previous section suggest a number of possible tests that can be
conducted using the observed properties of quasi-periodic oscillations
in X-ray binaries. In this section, we will discuss two such
methods. The first, relying only on the difference between the General
Relativistic and scalar-tensor orbital frequencies, will turn out to
be an as-yet inconclusive -- though rather general -- test. The
second, which relies on a proposed theoretical picture of QPOs,
generates tighter constraints.

\subsection{Limiting $\beta$ by orbital frequency alone}

In many QPO models, a strong peak in the Fourier spectrum of the
observed luminosity time-series is produced near the orbital frequency
of the inner edge of the Keplerian disk~\cite{p01}. In both General
Relativistic and scalar-tensor theories, the orbital frequency
increases monotonically with decreasing distance from the center of
the object.

Thus, assuming that the QPO frequencies come from some radius close to
but never less than that of the inner edge of the Keplerian flow where
the epicyclic frequency becomes imaginary, this suggests a simple null
test of scalar tensor theories. To wit: is the highest possible
orbital frequency one can produce around a scalar-tensor star -- for a
particular value of the parameter $\beta$ -- lower than any observed
QPO frequency? We may refine this question further by specifying a
reasonable (ADM) mass range we wish the scalar-tensor stars to have.

\begin{figure}
\includegraphics[width=8cm]{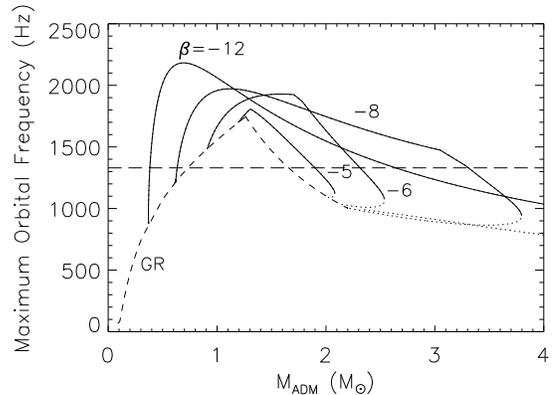}
\caption{The highest orbital frequency produced in scalar-tensor theories, as a function of mass, for equation of state UU. The long dashed line shows the current 1330 Hz lower limit on the maximum QPO frequency, the short dashed line represents the General Relativistic case, and the dotted lines indicate unstable branches.}
\label{5_1}
\end{figure}

The current lower limit on the maximum QPO frequency is 1330
Hz~\cite{sfkmk00}. We draw this lower limit on Fig.~\ref{5_1}, where
we show the highest orbital frequency for a variety of scalar-tensor
theories as a function of mass, for equation of state UU. The
nonintuitive property of scalar-tensor stars can immediately be seen:
making the parameter $\beta$ more negative increases the star radius
(see Fig.~\ref{3_3}) but also increases the orbital frequency at every
radius beyond the stellar surface (see Fig.~\ref{4_1}); this means
that a wider range of masses are able to produce the high maximum
frequency in scalar-tensor theory. There appears to be no simple way
of demonstrating the necessity of this result; all that can be said is
that the metric around a scalar-tensor star behaves very differently
from its Schwarzschild cousin.

\begin{figure}
\includegraphics[width=8cm]{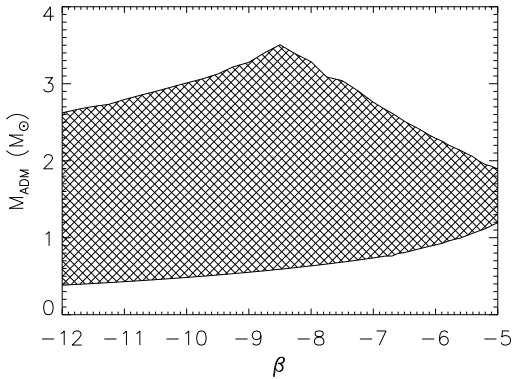}
\caption{Allowed range (indicated by crosshatching) for $M_{\textrm{\small{ADM}}}$, as a function of $\beta$, for EOS~UU, for the current best lower limit on the maximum QPO of 1330~Hz.}
\label{5_2}
\end{figure}

For completeness, we present this result in a complementary manner in
Fig.~\ref{5_2}. Here we show, as a function of the value of
$\beta$, the mass range for a variety of equations of state that can
reproduce the QPO lower limit of 1330 Hz. Thus this test, though ideal
because of its model independent nature, cannot rule out any part of
the parameter space.

\subsection{Limiting $\beta$ by joint analysis of orbital and epicyclic frequencies}

The high-frequency QPOs discussed in this paper often come in
pairs~\cite{v00} with the two peaks maintaining slowly changing
separations. The relativistic precession model~\cite{svm99} predicts a
simple relationship between, on the one hand, the positions and
separations of these QPO peaks and, on the other hand, the orbital and
epicyclic frequencies at the inner edge of the accretion disk of the
neutron star.

In this model, as the inner radius of the disk changes over time, the
QPO frequencies change to reflect the change in orbital and epicyclic
frequency. This relationship allows one to use measurements from a
single object to sample, not just a single radius in the neutron star
external spacetime, but a whole region of the orbital plane. This
results in much stronger constraints on the nature of strong-field
gravity while at the same time reducing the sensitivity to the choice
of equation of state for ultra-dense matter. As we will see, this
allows us, along with the assumption that the stars are
slowly-rotating, to rule out the scalar-tensor theory of gravity
contained in Eqs.~(\ref{lagrangian}) and~(\ref{a}) for a range of
the parameter $\beta$.

To derive constraints on the scalar-tensor theory we first solve the
equations of motion for particles in the external spacetime, for each
value of the parameter $\beta$ and $M_{\textrm{\small{ADM}}}$, to find
the Keplerian orbital frequency, $\omega_o$, and the epicyclic
frequency, $\kappa$, as a function of radius. The lower bound for the
radius is either that of the innermost stable orbit, or that at the
surface of the neutron star itself, whichever is larger. We then find
the relationship between $\omega_o$ and $\omega_o-\kappa$ themselves,
which according to the model correspond to the frequencies of the two
QPOs.

\begin{figure}
\includegraphics[angle=-90,width=8cm]{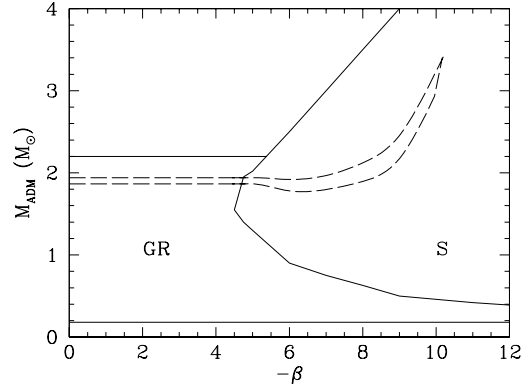}
\caption{Two-sigma confidence contours (dashed line), as a function of $M_{\textrm{\small{ADM}}}$, for a fit to the data of Ref.~\cite{mv00}, for a range of values of $\beta$ and EOS~UU. The regions bounded by solid lines show the region of standard GR behaviour (``GR'' region) and the parameter space in which stars undergo spontaneous scalarization (``S'' region.)}
\label{6_1}
\end{figure}

We then fit this relationship to the observed QPO frequency pairs, and
derive a reduced chi-squared value for each point in
$(M_{\textrm{\small{ADM}}}, \beta)$ space. In Fig.~\ref{6_1}, we plot
the two-sigma confidence contour around the minimum $\chi^2$ value
obtained for the GR case. As can be seen, the onset of scalarization
alters the range of allowed masses, and for values of
$\beta\lesssim-10.25$, scalarization produces neutron stars with
spacetimes sufficiently different from the GR case as to be ruled out
altogether.

We emphasize that this test has assumed that the rotation of the
neutron star itself has a small (less than order unity) effect on the
external spacetime. While we have not constructed and solved the
equations for the scalar-tensor theory in the case of a rotating
matter source, we believe that the corrections to the
orbital-epicyclic frequency relationship which governs this test will
be on the order of $20\%$.

We expect that combining observations from different systems, and
requiring that $\beta$ be held fixed across all observations, will
provide additional confirmation of this result. Because of the
uncertainty introduced by the consideration of nonzero stellar angular
momentum, we cannot put a precise lower bound on the value of the
parameter $\beta$, but given the magnitude of corrections found for
rotation in the GR case, the lower bound is presumably not far from
$\beta\sim-10$.

The analysis of this section has relied on a particular model of QPO
production. Other QPO models, since they also invoke processes at the
high spacetime curvatures near the surface of the star, will lead to
similar kinds of constraints, of varying severity, for scalar-tensor
models.

\section{Discussion}

We have investigated two different methods for constraining the nature
of strong-field gravity using observations of QPOs. As an example of a
theory of gravity that deviates from GR, we have used a particular
scalar-tensor theory, parametrized by a single real number,
$\beta$. The class of theories we have investigated has gained a great
deal of attention because of its invisibility to traditional,
weak-field tests.

We have investigated the nature of test particle orbits in the
external spacetime of the neutron star, and, while uncovering some
counterintuitive behavior, have confirmed for test particle orbits
what has already been known for other observables of compact objects
in this scalar-tensor theory: while large (order unity or greater)
deviations from General Relativity are observed within the first few
hundred Schwarzschild radii of the neutron star, the effects of
introducing a scalar field rapidly disappear at the larger distances
(and far lower curvatures) probed by ordinary weak-field tests and
require observations of much greater precision to detect.

The first method we have investigated, while not relying on a
particular theory of QPO production, does not produce strong
constraints on the value of the parameter $\beta$. The second method,
which relies upon the relativistic precession model of QPO production,
is more powerful in its ability to separate the predictions of
General Relativity from those of scalar-tensor theory. The power of
this second method derives in part from the fact that we are able to
sample a range of radii around the neutron star, and not just
a single radius.

In a previous paper~\cite{dp2}, we investigated the use of
gravitationally redshifted atomic lines in testing strong-field
gravity, and showed that the nature of
gravity in the strong-field is less well known than the equation of
state for ultra-dense matter. Further constraining this equation of
state has often been taken to be the goal of
theoretical studies of neutron star properties.

We find, however, that a second, promising avenue is opening up
for investigations that attempt to constrain the nature of gravity
itself. In addition to the tests presented here and in previous
work, additional ways of testing strong-field gravity can be
performed that take advantage of other aspects of the rapid
observational and theoretical progress that has been made in recent
years in the study of compact objects in the X-ray and soft-gamma ray
bands. Another such test based on the Eddington-limited luminosities of 
radius-expansion bursts will be reported~\cite{dp4}.

While solar system and binary pulsar tests
observe with a much higher degree of precision than can be expected
from the investigation of high-energy processes near the surface of
compact objects, the large spacetime curvatures found in the latter
systems can, in some circumstances, magnify the expected deviations and
produce constraints on the nature of GR of similar severity. In such
cases, weak field tests and strong-field tests described in this
paper and others may be considered complementary.

\end{document}